\documentclass{article}

\usepackage{arxiv}
\usepackage{amsmath} 
\usepackage{array}
\usepackage[utf8]{inputenc} 
\usepackage[T1]{fontenc}    
\usepackage{hyperref}       
\usepackage{url}            
\usepackage{booktabs}       
\usepackage{amsfonts}       
\usepackage{nicefrac}       
\usepackage{microtype}      
\usepackage{lipsum}		
\usepackage{graphicx}
\usepackage{natbib}
\usepackage{doi}

\title{Wavefield Finite Time Focusing with Reduced Spatial Exposure}

\author{
  Giovanni Angelo Meles\\
    Faculty of Civil Engineering and Geosciences\\
      Delft University of Technology \\
      The Netherlands\\
   \And
  Joost van der Neut \\
    Faculty of Applied Sciences\\
      Delft University of Technology \\
      The Netherlands\\
  \And
  Koen W. A. van Dongen \\
    Faculty of Applied Sciences\\
      Delft University of Technology \\
      The Netherlands\\
  \And
  Kees Wapenaar\\
    Faculty of Civil Engineering and Geosciences\\
      Delft University of Technology \\
      The Netherlands\\
}

\renewcommand{\headeright}{Technical Report}
\renewcommand{\undertitle}{Technical Report}
\renewcommand{\shorttitle}{Wavefield finite time focusing with reduced spatial exposure}

\hypersetup{
pdftitle={A template for the arxiv style},
pdfsubject={q-bio.NC, q-bio.QM},
pdfauthor={David S.~Hippocampus, Elias D.~Striatum},
pdfkeywords={First keyword, Second keyword, More},
}
\date{}
\begin{document}
\maketitle

\begin{abstract}
Wavefield focusing is often achieved by Time-Reversal Mirrors, where wavefields emitted by
a source located at the focal point are evaluated at a closed boundary and sent back, after Time-Reversal, into the medium from that boundary.
Mathematically, Time-Reversal Mirrors are derived from closed-boundary integral representations of reciprocity theorems.
In heterogeneous media, Time-Reversal Focusing theoretically involves in- and output signals that are infinite in time and the resulting  waves propagate through the entire medium. 
Recently, integral representations have been derived for single-sided wavefield focusing. 
Although the required input signals for this approach are finite in time, the output signals are not and, similar to Time-Reversal Mirroring, the resulting waves propagate through the entire medium.
Here, an alternative solution for double-sided wavefield focusing is derived. 
This solution is based on an integral representation where in- and output signals are finite in time, and where the energy of the waves propagating in the layer embedding the focal point is smaller than 
with Time-Reversal Focusing. 
We explore the potential of the proposed method with numerical experiments involving a head model consisting of a skull enclosing a brain.
\end{abstract}

\addtocounter{page}{2}

 \setlength{\parindent}{5ex}

\section{INTRODUCTION}

With Time-Reversal Mirrors, wavefields can be focused at a specified focal point in an arbitrary heterogeneous medium \cite{fink93}. To realize such a mirror, wavefields from
a source at the focal point are evaluated at a closed boundary and sent back, after Time-Reversal, into the medium from that boundary. As can be demonstrated from Green's theorem, this procedure leads to 
a solution of the homogeneous wave equation, consisting of an acausal wavefield that focuses at the focal point and a causal wavefield,
propagating from the focal point through the entire medium to the boundary \cite{wapenaar06,fink2006}. 
Applications can be found in various areas. In medical acoustics, Time-Reversal Mirroring has been applied for kidney stone and tumor ablation \cite{thomas96,aubry07}. 
The Time-Reversal concept is also a key ingredient for various source localization \cite{catheline07,li16} and reflection imaging \cite{mcmechan83,oristaglio89} algorithms. 
Assuming that the medium is lossless and sufficiently heterogeneous, both the acausal wavefield that propagates towards 
the focal point and the causal wavefield that propagates through the medium to the boundary are unbounded in time.\\

Recently, it was shown that wavefields in one-dimensional media can also be focused from a single open-boundary by solving the Marchenko equation \cite{rose02}, being a familiar
result from inverse scattering theory \cite{burridge80}. In this case a different focusing condition is achieved  \cite{wapenaar14c}, and when the solution of the Marchenko equation 
is emitted into the medium from a single open-boundary, a focus emerges at the focal point, followed by a causal Green's function that propagates from the focal point through the entire medium to the boundary \cite{broggini12}. 
This result can be extended to three-dimensional wave propagation \cite{wapenaar14} and various focusing conditions \cite{meles18} and has seen various applications in exploration geophysics, such as reflection imaging \cite{ravasi16} 
and acoustic holography \cite{wapenaar16}. Although the focusing function is finite in time, the Green's function that emerges after wavefield focusing has infinite duration. 
In this paper, it will be discussed how to craft a focusing wavefield that, once injected in the medium from two open-boundaries, propagates to a specified focal point in finite time,
without being followed by any Green's
function. It will also be discussed how this focusing method theoretically reduces wavefield propagation in the layer which embeds the focal point.
Numerical tests involving a complex model will show that wavefield propagation is largely reduced in the layer embedding the focal point despite the fact that exact focusing functions cannot be retrieved.

\section{THEORY}

Coordinates in three-dimensional space are defined as ${\bf{x}}=\left(x_1,x_2,x_3\right)$, and $t$ denotes time. Although the derived theory can be modified for various types 
of wave phenomena, acoustic wave propagation is considered. The medium is lossless and characterized by propagation velocity $c \left({\bf{x}}\right)$ and mass density $\rho \left({\bf{x}}\right)$. It is assumed that 
these properties are independent of time. The acoustic pressure wavefield is expressed as $p \left( {\bf{x}}, t \right)$. 
For simplicity all derivations are carried out in the frequency domain, and the temporal Fourier transform of $p \left( {\bf{x}}, t \right)$ is
defined by $p \left( {\bf{x}}, \omega \right) =\int_{-\infty}^\infty
p \left( {\bf{x}}, t \right)\exp{\left(i \omega t\right)} dt$, where $\omega$ is the angular frequency. All wavefields obey the wave equation, which is defined in the frequency domain as

\begin{equation} \label{eq:T1}
\partial_i
\left(
\frac{1}{\rho(\bf{x})}
\partial_i
{{p}\left( {\bf{x}}, \omega \right)}
\right)
+
\frac{
\omega^{2}}
{\rho(\bf{x}) \mbox{c$^2$}(\bf{x}) }
{{p}\left( {\bf{x}}, \omega \right)}
=
i \omega
{{q}\left( {\bf{x}}, \omega \right)}
,\end{equation}
with $\partial_i $ standing for the spatial derivative $\frac{\partial}{\partial x_i}$ , where
 $i$ takes the values $1,2$ and $3$. Einstein's summation
convention is applied, meaning that summation is carried out over repeated indeces. Note that the source function ${{q}\left( {\bf{x}}, \omega \right)}$, 
standing for volume-injection rate density,
is scaled by $i \omega$. Since the wave equation is often defined without this scaling factor elsewhere in the literature, the wavefields that appear in this paper should be divided
with $\left(i \omega \right)$ to be consistent with that literature. The Green's function ${{G}\left( {\bf{x}}, {\bf{x}}_S, \omega \right)}$ is defined as the solution of the wave
equation for ${{q}\left( {\bf{x}}, \omega \right)}=\delta \left({\bf{x}}-{\bf{x}}_S\right)$, where ${\bf{x}}_S$ is the source location.\\

\begin{figure}
  \centering
   \includegraphics[width=0.85\textwidth]{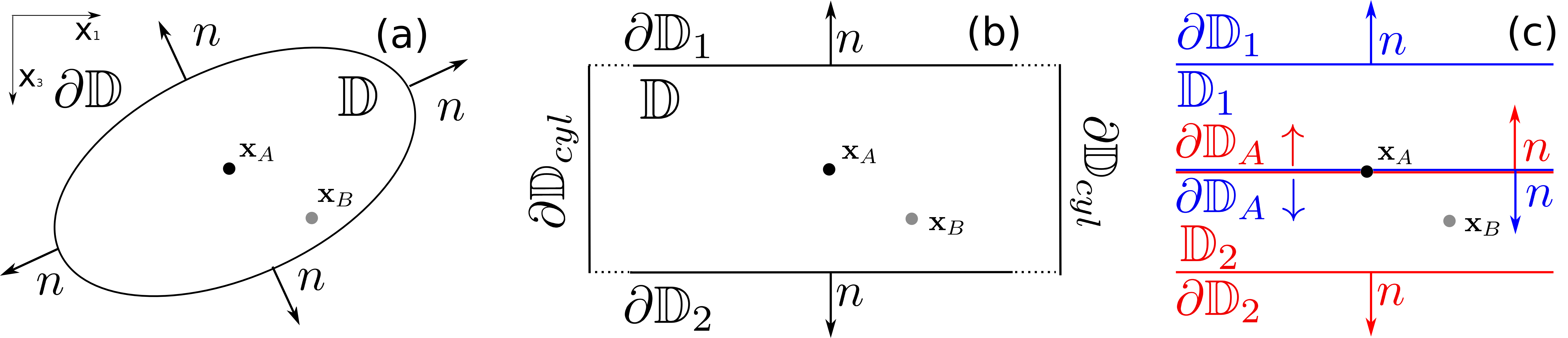}
\caption{(Color online) (a) Cross-section of the configuration in the $\left(x_1,x_3\right)$-plane for Eq. (\ref{eq:T2c}). Volume ${\mathbb{D}}$ is enclosed by $\partial{\mathbb{D}}$ (solid line) with outward-pointing normal
vectors $n$. 
(b) Cross-section of the configuration  for Eq. (\ref{eq:T2l}). Volume ${\mathbb{D}}$ is enclosed by $\partial \mathbb{D}_1 \cup \partial\mathbb{D}_2 \cup \partial\mathbb{D}_{cyl}$ (solid black lines). 
(c) Cross-section of the configuration  for Eq. (\ref{eq:T9}) 
Volume ${\mathbb{D}}$ is splitted into ${\mathbb{D}}_1$ and  $ {\mathbb{D}}_2$, surrounded by  
$\partial {\mathbb{D}}_1 \cup \partial {\mathbb{D}}_A$ (blue line) and $\partial {\mathbb{D}}_2 \cup \partial {\mathbb{D}}_A$ (red line), respectively. 
Note that the normals $n$ relative to ${\partial \mathbb{D}}_1 \cup \partial {\mathbb{D}}_A$ and ${\partial \mathbb{D}}_2 \cup \partial {\mathbb{D}}_A$ across $\partial \mathbb{D}_A$
are antiparallel. The focal point is at $\bf{x}_A \in \partial{\mathbb{D}}_A$.
\label{fig:J1}}
\end{figure}


It has been shown how the real part of the Green's function with a source at ${\bf{x}}_A$ and a receiver at ${\bf{x}}_B$ can be expressed by integrating a specific combination of observations from 
 sources at ${\bf{x}}_A$ and ${\bf{x}}_B$ over \textit{any} boundary $\partial {\mathbb{D}}$ 
that encloses volume ${\mathbb{D}}$, where ${\bf{x}}_A \in {\mathbb{D}}$ and ${\bf{x}}_B \in {\mathbb{D}}$ (Fig. \ref{fig:J1}a):
 
\begin{equation} \label{eq:T2c}
\begin{split}
&2 \Re \lbrace{ G \left( {\bf{x}}_B, {\bf{x}}_A ; \omega \right)}\rbrace 
\\
&=
\oint_{\partial \mathbb{D}}
d^2{\bf{x}}
\frac{1}{j \omega \rho(\bf{x})}
\left(
G\left( {\bf{x}}, {\bf{x}}_B, \omega \right)
n_i \partial_i G^* \left( {\bf{x}}, {\bf{x}}_A, \omega \right)
-
G^* \left( {\bf{x}}, {\bf{x}}_A, \omega \right)
n_i \partial_i
G\left( {\bf{x}}, {\bf{x}}_B, \omega \right)
\right),
\end{split}
\end{equation}
where $n_i$ is the outward pointing normal of ${\partial \mathbb{D}}$ and superscript $*$ denotes complex conjugation. We call Eq. (\ref{eq:T2c}) a representation of
the Green's function $G( \bf{x}_B, \bf{x}_A ; \omega)$. In Time-Reversed acoustics, 
observations from a source at ${\bf{x}}_A$ are 
reversed in time and injected into the medium at ${\partial \mathbb{D}}$.
The complex-conjugate Green's function $G^*(x,x_A,\omega)$ stands for the Fourier transform of the time-reversed observations.
Equation (\ref{eq:T2c}) can thus be interpreted as if the injected field were propagated forward in time to any 
location ${\bf{x}}_B$ by the Green's function $G\left( {\bf{x}}_B, {\bf{x}}, \omega \right)$, which is equal to $G\left( {\bf{x}}, {\bf{x}}_B, \omega \right)$ through source-receiver 
reciprocity \cite{fokkema93}. As can be learned from Eq. (\ref{eq:T2c}), this procedure yields for any location ${\bf{x}}_B$ the real part of the Green's function $G \left( {\bf{x}}_B, {\bf{x}}_A ; \omega \right)$, which can be
interpreted as the Fourier transform of the superposition of an acausal Green's function, focusing at ${\bf{x}} = {\bf{x}}_A$, and a causal Green's function that propagates from $x_A$ through 
the entire medium to ${\partial \mathbb{D}}$. Since the 
source functions of this acausal and causal Green's function cancel each other, their superposition satisfies the homogeneous wave equation (i.e. Eq. (\ref{eq:T1}) for ${{q}\left( {\bf{x}}, \omega \right)}=0$). Note that
this homogeneous wave equation is valid also for heterogeneous media.
Note also that Time-Reversed acoustics results in a wavefield that at time $t=0$  is non-zero just at the focal point \cite{wapenaar14b},
but it poses no constraints on the wavefield at other times.\\

We also consider a peculiar closed boundary $\partial \mathbb{D}= \partial \mathbb{D}_1 \cup \partial \mathbb{D}_2 \cup \partial \mathbb{D}_{cyl}$, where $\partial \mathbb{D}_1$ and
$\partial \mathbb{D}_2$ are horizontal boundaries connected by a cylindrical surface $\partial \mathbb{D}_{cyl}$ with infinite radius  (Fig. \ref{fig:J1}b).
For this configuration, the contribution of the integral in Eq. (\ref{eq:T2c}) over $\partial \mathbb{D}_{cyl}$ vanishes and the following representation holds \cite{wapenaar16}:

\begin{equation} \label{eq:T2l}
\begin{split}
&2 \Re \lbrace{ G \left( {\bf{x}}_B, {\bf{x}}_A ; \omega \right)}\rbrace
\\
&=
\int_{\partial \mathbb{D}_1 \cup \partial \mathbb{D}_2}
d^2{\bf{x}}
\frac{1}{j \omega \rho(\bf{x})}
\left(
G\left( {\bf{x}}, {\bf{x}}_B, \omega \right)
n_3 \partial_3 G^* \left( {\bf{x}}, {\bf{x}}_A, \omega \right)
-
G^* \left( {\bf{x}}, {\bf{x}}_A, \omega \right)
n_3 \partial_3
G\left( {\bf{x}}, {\bf{x}}_B, \omega \right)
\right).
\end{split}
\end{equation}

In addition to standard Time-Reversed acoustics, interesting focusing wavefields can be derived also by using focusing functions, which have
recently been introduced to denote the solutions of the multidimensional Marchenko equation \cite{wapenaar14}. 
In this derivation, the same horizontal boundaries $\partial \mathbb{D}_1$ and $\partial \mathbb{D}_2$ as in Eq. (\ref{eq:T2l}) are used, but an
additional auxiliary boundary $\partial{\mathbb{D}}_A$ is introduced. Here, $\partial{\mathbb{D}}_A$ is a horizontal plane inside ${\mathbb{D}}$ that intersects with the 
focal point ${\bf{x}}_A=\left( x_{1,A}, x_{2,A},x_{3,A}\right)$, so that volume ${\mathbb{D}}$ is divided into a subvolume ${\mathbb{D}}_1$, located above $\partial{\mathbb{D}}_A$, and a
subvolume ${\mathbb{D}}_2$, located below $\partial{\mathbb{D}}_A$ (Fig. \ref{fig:J1}c). 
Note that the normals along $\partial{\mathbb{D}_A}$ associated with subvolumes ${\mathbb{D}}_1$ and ${\mathbb{D}}_2$ are antiparallel (Fig. \ref{fig:J1}c).\\

We deduce new sets of representation theorems for volumes ${\mathbb{D}}_1$ and ${\mathbb{D}}_2$. First of all, a reciprocity theorem of 
the convolution type \cite{fokkema93} associated with volume ${\mathbb{D}}_1$ is introduced:

\begin{equation} \label{eq:T3v1}
\int_{\mathbb{D}_1}
d^3{\bf{x}}
\left( {{p}}_A {{q}}_{B}-{{p}}_{B} {{q}}_A \right) =\int_{\partial \mathbb{D}_1}d^2{\bf{x}} \frac{1}{j \omega \rho}\left(  {p}_{B} n_3 \partial_3 {p}_{A}  - {p}_A n_3 \partial_3 {p}_{B} \right)
- \int_{\partial \mathbb{D}_A}d^2{\bf{x}}\frac{2}{j \omega \rho}\left( {{p}}^{+}_A \partial_3 {p}^{-}_{B} + {p}^{-}_A \partial_3 {p}^{+}_{B}\right).
\end{equation}

Subscripts $A$ and $B$ indicate two states. The integral over ${\partial \mathbb{D}}_A$ has been modified 
by using fundamental properties \cite{fishman93} of the (Helmholtz) operator in Eq. (\ref{eq:T2c}), where the wavefields have been decomposed into downgoing (indicated by superscript $+$) and upgoing 
(indicated by superscript $-$) constituents. In addition, the field has been normalized such that $p=p^++p^-$. Similarly, a reciprocity theorem of the correlation type \cite{bojarski83} can be modified as

\begin{equation} \label{eq:T4v1}
\int_{\mathbb{D}_1}
d^3{\bf{x}}
\left({{p}}_A^* {{q}}_{B} +{{p}}_{B} {{q}}_A^*  \right)=\int_{\partial \mathbb{D}_1}d^2{\bf{x}}\frac{1}{j \omega \rho}\left({p}_{B} n_3 \partial_3 {p}_{A}^{*}-{p}_A^{*}  n_3 \partial_3 {p}_{B}\right) 
-\int_{\partial \mathbb{D}_A}d^2{\bf{x}}\frac{2}{j \omega \rho}\left( {{p}}^{+*}_A \partial_3 {p}^{+}_{B} + {p}^{-*}_A  \partial_3 {p}^{-}_{B}\right).
\end{equation}

Two representations will be derived for subvolume $\mathbb{D}_1$. In both representations, state $A$ is source-free (${{q}}_A=0$). The medium properties in this state are identical to the physical
properties $c\left({\bf{x}}\right)$ and $\rho\left({\bf{x}}\right)$ within $\mathbb{D}_1$, and can be arbitrarily set below $\partial \mathbb{D}_A$ \cite{wapenaar14}.
Here, the properties of the medium are chosen such that the halfspace below $\partial \mathbb{D}_A$ is non-scattering.
A particular solution of the source-free wave equation will be substituted in this state, which is referred to as focusing function $p_A=f_1(\textbf{x},\textbf{x}_A,\omega)$, 
where  $\textbf{x}_A$ is the focal point and $\textbf{x}$ is a variable coordinate inside the domain $\mathbb{D}$  \cite{wapenaar14}. This
focusing function is subject to a different focusing condition than what is achieved by Time-Reversed acoustics. 
In this paper, the condition is defined 
as ${{f}}_1^{+} \left( {\bf{x}}, {\bf{x}}_A ;\omega \right) |_{{\bf{x}} \in {\partial \mathbb{D}_A}}=\delta \left( {\bf{x}}_{H} - {\bf{x}}_{H,A} \right)$, where ${\bf{x}}_H=\left(x_1, x_2\right)$ is a point in the focal
plane, 
while ${{f}}_1^{-} \left( {\bf{x}}, {\bf{x}}_A ;\omega \right) |_{{\bf{x}} \in {\partial \mathbb{D}_A}}$ vanishes.

The first condition states that the downgoing part of the focusing function focuses at $\textbf{x}_A$ not followed by any other event.
This is achieved by cancelling any further down-going wave via destructive interference with propagation of the coda of the focusing function (see \cite{wapenaar14} for more details).
After having focused, this downgoing function continues its propagation into the 
lower half-space. Since the lower half-space was chosen to be scattering-free, the upgoing part of the focusing function at $\partial \mathbb{D}_A$ is zero.
Note that this condition does not pose any constraint on the wavefield at time $t=0$ away from the focal plane $\partial \mathbb{D}_A$.
In state $B$, the medium properties are equivalent to the physical medium, where an 
impulsive source is located at ${\bf{x}}_B \in \mathbb{D}$, yielding ${{q}}_{B}= \delta{ \left( {\bf{x}} - {\bf{x}}_B \right)}$ and ${{p}}_{B}={{G}} \left( {\bf{x}}, {\bf{x}}_B ; \omega \right)$. 
Substituting these quantities into Eqs. (\ref{eq:T3v1}) and (\ref{eq:T4v1}) brings

\begin{equation} \label{eq:T5v1}
\begin{split}
\theta \left( x_{3,A} - x_{3,B} \right)
& {f}_1 \left( {\bf{x}}_B, {\bf{x}}_A ; \omega \right)
+
\frac{2}{j \omega \rho(\bf{x}_A)} \partial_3 G^- \left({\bf{x}}_A, {\bf{x}}_B, \omega\right)
 =
\int_{\partial \mathbb{D}_1}
d^2{\bf{x}}
\frac{1}{j \omega \rho \left({\bf{x}}\right) }
\times
\\
& \left(
G \left( {\bf{x}}, {\bf{x}}_B, \omega \right)
n_3 \partial_3 
f_1 \left( {\bf{x}}, {\bf{x}}_A, \omega \right)
-
f_1 \left( {\bf{x}}, {\bf{x}}_A, \omega \right)
n_3 \partial_3 
G \left( {\bf{x}}, {\bf{x}}_B, \omega \right)
\right)
,
\end{split}
\end{equation}
and
\begin{equation} \label{eq:T6v1}
\begin{split}
\theta \left( x_{3,A} - x_{3,B} \right)
& {f}_1^* \left( {\bf{x}}_B, {\bf{x}}_A ; \omega \right)
+
\frac{2}{j \omega \rho(\bf{x}_A)} \partial_3 G^+ \left({\bf{x}}_A, {\bf{x}}_B, \omega\right)
 =
\int_{\partial \mathbb{D}_1}
d^2{\bf{x}}
\frac{1}{j \omega \rho \left({\bf{x}}\right) }
\times
\\
& \left(
G \left( {\bf{x}}, {\bf{x}}_B, \omega \right)
n_3 \partial_3 
f_1^* \left( {\bf{x}}, {\bf{x}}_A, \omega \right)
-
f_1^* \left( {\bf{x}}, {\bf{x}}_A, \omega \right)
n_3 \partial_3 
G \left( {\bf{x}}, {\bf{x}}_B, \omega \right)
\right),
\end{split}
\end{equation}
 where $\theta \left( x_{3}\right)$ is a Heaviside function, with $\theta \left( x_{3}\right)=0$ for $x_{3}<0$, $\theta \left( x_{3}\right)=\frac{1}{2}$ for
$x_{3}=0$ and $\theta\left( x_{3}\right)=1$ for $x_{3}>0$.\\

Convolution and correlation reciprocity theorems associated with volume ${\mathbb{D}}_2$ are also introduced:

\begin{equation} \label{eq:T3v2}
\int_{\mathbb{D}_2}
d^3{\bf{x}}
\left( {{p}}_A {{q}}_{B}-{{p}}_{B} {{q}}_A \right) =\int_{\partial \mathbb{D}_2}d^2{\bf{x}} \frac{1}{j \omega \rho}\left(  {p}_{B} n_3 \partial_3 {p}_{A}  - {p}_A n_3 \partial_3 {p}_{B} \right)
+ \int_{\partial \mathbb{D}_A}d^2{\bf{x}}\frac{2}{j \omega \rho}\left( {{p}}^{+}_A \partial_3 {p}^{-}_{B} + {p}^{-}_A \partial_3 {p}^{+}_{B}\right),
\end{equation}

\begin{equation} \label{eq:T4v2}
\int_{\mathbb{D}_2}
d^3{\bf{x}}
\left({{p}}_A^* {{q}}_{B} +{{p}}_{B} {{q}}_A^*  \right)=\int_{\partial \mathbb{D}_2}d^2{\bf{x}}\frac{1}{j \omega \rho}\left({p}_{B} n_3 \partial_3 {p}_{A}^{*}-{p}_A^{*}  n_3 \partial_3 {p}_{B}\right) 
+\int_{\partial \mathbb{D}_A}d^2{\bf{x}}\frac{2}{j \omega \rho}\left( {{p}}^{+*}_A \partial_3 {p}^{+}_{B} + {p}^{-*}_A  \partial_3 {p}^{-}_{B}\right).
\end{equation}

Two representations can be similarly derived for subvolume $\mathbb{D}_2$. For both representations, state $A$ is source-free (${{q}}_A=0$), with medium properties as in the physical state 
in $\mathbb{D}_2$ and a non-scattering halfspace above $\partial \mathbb{D}_A$. Focusing function $p_A=f_2(\textbf{x},\textbf{x}_A,\omega)$ will be substituted, being a solution of the source-free 
wave equation, with the focusing condition 
${{f}}_2^{-} \left( {\bf{x}}, {\bf{x}}_A ;\omega \right) |_{{\bf{x}} \in {\partial \mathbb{D}_A}}=\delta \left( {\bf{x}}_{H} - {\bf{x}}_{H,A} \right)$, while 
${{f}}_2^{+} \left( {\bf{x}}, {\bf{x}}_A ;\omega \right) |_{{\bf{x}} \in {\partial \mathbb{D}_A}}$ 
vanishes. In state $B$, conditions are the same as in the derivation of the 
previous representations. Substituting these quantities into Eq. (\ref{eq:T3v2}) and Eq. (\ref{eq:T4v2}) yields

\begin{equation} \label{eq:T5v2}
\begin{split}
\theta \left( x_{3,B} - x_{3,A} \right)
& {f}_2 \left( {\bf{x}}_B, {\bf{x}}_A ; \omega \right)
-
\frac{2}{j \omega \rho(\bf{x}_A)} \partial_3 G^+ \left({\bf{x}}_A, {\bf{x}}_B, \omega\right)
 =
\int_{\partial \mathbb{D}_2}
d^2{\bf{x}}
\frac{1}{j \omega \rho \left({\bf{x}}\right) }
\times
\\
& \left(
G \left( {\bf{x}}, {\bf{x}}_B, \omega \right)
n_3 \partial_3
f_2 \left( {\bf{x}}, {\bf{x}}_A, \omega \right)
-
f_2 \left( {\bf{x}}, {\bf{x}}_A, \omega \right)
n_3 \partial_3 
G \left( {\bf{x}}, {\bf{x}}_B, \omega \right)
\right),
\end{split}
\end{equation}
and
\begin{equation} \label{eq:T6v2}
\begin{split}
\theta \left( x_{3,B} - x_{3,A} \right)
& {f}_2^* \left( {\bf{x}}_B, {\bf{x}}_A ; \omega \right)
-
\frac{2}{j \omega \rho(\bf{x}_A)} \partial_3 G^- \left({\bf{x}}_A, {\bf{x}}_B, \omega\right)
 =
\int_{\partial \mathbb{D}_2}
d^2{\bf{x}}
\frac{1}{j \omega \rho \left({\bf{x}}\right) }
\times
\\
& \left(
G \left( {\bf{x}}, {\bf{x}}_B, \omega \right)
n_3 \partial_3
f_2^* \left( {\bf{x}}, {\bf{x}}_A, \omega \right)
-
f_2^* \left( {\bf{x}}, {\bf{x}}_A, \omega \right)
n_3 \partial_3 
G \left( {\bf{x}}, {\bf{x}}_B, \omega \right)
\right).
\end{split}
\end{equation}

In the following we discuss two focusing strategies based on the focusing functions introduced in Eqs. (\ref{eq:T5v1})-(\ref{eq:T6v1}) and (\ref{eq:T5v2})-(\ref{eq:T6v2}).\\

Standard (double-sided) Marchenko Focusing can be achieved by injecting $f_1$ and $f_2$ from $\partial \mathbb{D}_1$ and $\partial \mathbb{D}_2$, respectively. The corresponding wavefields 
propagate from $\partial \mathbb{D}_1$ and $\partial \mathbb{D}_2$ to the focal point, subsequently generating scattering events in  $\mathbb{D}_2$ and $\mathbb{D}_1$. 
Note that focusing functions $f_1$ and $f_2$ are defined in reference states involving non-scattering media below or above $\partial \mathbb{D}_A$ \cite{wapenaar14}, 
but in this physical experiment they are injected in the actual medium, thus generating scattering events below or above $\partial \mathbb{D}_A$. 
These scattered wavefields eventually interfere with the focal plane.
Standard (double-sided) Marchenko Focusing can be mathematically expressed by the summation of Eqs. (\ref{eq:T5v1}) and (\ref{eq:T5v2}):

\begin{equation} \label{eq:T7v1}
\begin{split}
\theta \left( x_{3,A} - x_{3,B} \right){f}_1 \left( {\bf{x}}_A, {\bf{x}}_B ; \omega \right) +  \theta \left( x_{3,B} - x_{3,A} \right) {f}_2 \left( {\bf{x}}_B, {\bf{x}}_A ; \omega \right)+\\
\frac{2}{j \omega \rho(\bf{x}_A)} \partial_3 G^+ \left({\bf{x}}_A, {\bf{x}}_B, \omega\right) - \frac{2}{j \omega \rho(\bf{x}_A)} \partial_3 G^+ \left({\bf{x}}_A, {\bf{x}}_B, \omega\right)
 = \\
 \int_{\partial \mathbb{D}_1}
d^2{\bf{x}}
\frac{1}{j \omega \rho \left({\bf{x}}\right) }
\times
 \left(
G \left( {\bf{x}}, {\bf{x}}_B, \omega \right)
n_3 \partial_3
f_1 \left( {\bf{x}}, {\bf{x}}_A, \omega \right)
-
f_1 \left( {\bf{x}}, {\bf{x}}_A, \omega \right)
n_3 \partial_3 
G \left( {\bf{x}}, {\bf{x}}_B, \omega \right)
\right) +
 \\
\int_{\partial \mathbb{D}_2}
d^2{\bf{x}}
\frac{1}{j \omega \rho \left({\bf{x}}\right) }
\times
 \left(
G \left( {\bf{x}}, {\bf{x}}_B, \omega \right)
n_3 \partial_3
f_2 \left( {\bf{x}}, {\bf{x}}_A, \omega \right)
-
f_2 \left( {\bf{x}}, {\bf{x}}_A, \omega \right)
n_3 \partial_3 
G \left( {\bf{x}}, {\bf{x}}_B, \omega \right)
\right),
\end{split}
\end{equation}

An additional focusing strategy can be derived by further inspection and manipulation of Eqs. (\ref{eq:T5v1})-(\ref{eq:T6v1}) and (\ref{eq:T5v2})-(\ref{eq:T6v2}).  
The different orientation of the normals along $\partial{\mathbb{D}_A}$ when associated with subvolumes ${\mathbb{D}}_1$ or ${\mathbb{D}}_2$ results in opposite signs of the Green's functions terms
in the left-hand sides of Eqs. (\ref{eq:T5v1})-(\ref{eq:T6v1}) and (\ref{eq:T5v2})-(\ref{eq:T6v2}), respectively. Therefore, when Eq. (\ref{eq:T5v1}), (\ref{eq:T6v1}), (\ref{eq:T5v2}) and (\ref{eq:T6v2}) are added together, 
these Green's functions terms cancel out and it follows that:

\begin{equation} \label{eq:T9}
\begin{split}
&2\Re
\lbrace
{f} \left( {\bf{x}}_B, {\bf{x}}_A ; \omega \right)
\rbrace
\\
&=
\int_{\partial \mathbb{D}_1 \cup \mathbb{D}_2}
d^2{\bf{x}}
\frac{1}{j \omega \rho \left({\bf{x}}\right) }
 \left(
G \left( {\bf{x}}, {\bf{x}}_B, \omega \right)
n_3 \partial_3 
2\Re
\lbrace
f \left( {\bf{x}}, {\bf{x}}_A, \omega \right)
\rbrace
-
2\Re
\lbrace
f \left( {\bf{x}}, {\bf{x}}_A, \omega \right)
\rbrace
n_3 \partial_3 
G \left( {\bf{x}}, {\bf{x}}_B, \omega \right)
\right)
,
\end{split}
\end{equation}
where

\begin{equation} \label{eq:T10}
\begin{split}
{f} \left( {\bf{x}}, {\bf{x}}_A ; \omega \right)
=
\theta \left( {\bf{x}}_{3,A} - {\bf{x}}_{3} \right)
{f}_1 \left( {\bf{x}}, {\bf{x}}_A ; \omega \right)
+
\theta \left( {\bf{x}}_{3} - {\bf{x}}_{3,A} \right)
{f}_2 \left( {\bf{x}}, {\bf{x}}_A ; \omega \right)
.
\end{split}
\end{equation}

Akin to Eqs. (\ref{eq:T2c}) and (\ref{eq:T7v1}), this result can be used for wavefield focusing. By injecting the real part of the wavefield $f\left( {\bf{x}}, {\bf{x}}_A ; \omega \right)$, as 
defined by Eq. (\ref{eq:T10}), into the medium at 
boundaries ${\partial \mathbb{D}_1}$ and ${\partial \mathbb{D}_2}$, one can reconstruct this wavefield throughout the volume, as shown by Eq. (\ref{eq:T9}). Due to the intrinsic properties of 
focusing functions, i.e. the destructive interference of the codas with up- and down-going reflections, any scattering event is confined within a spatial-temporal window
defined by the propagation of the initial component of the focusing function (for more details see \cite{wapenaar14}).
As a consequence, the wavefield in Eq. (\ref{eq:T9}) propagates towards the focal
point in finite time and back to the surface in finite time again.

Moreover, due to the focusing properties of $f_1$ and $f_2$, the wavefield $f$ theoretically 
interacts with the focal plane $\partial{\mathbb{D}}_A$ only at  ${\bf{x}} = {\bf{x}}_{H,A}$ at $t=0$.
We refer to the focusing achieved by Eq. (\ref{eq:T9}) as 'Finite Time Focusing with reduced spatial exposure', which we will often abbreviate as 'Finite Time Focusing'.\\

\section{NUMERICAL EXAMPLES}

  \begin{figure}
  \centering
   \includegraphics[width=0.65\textwidth]{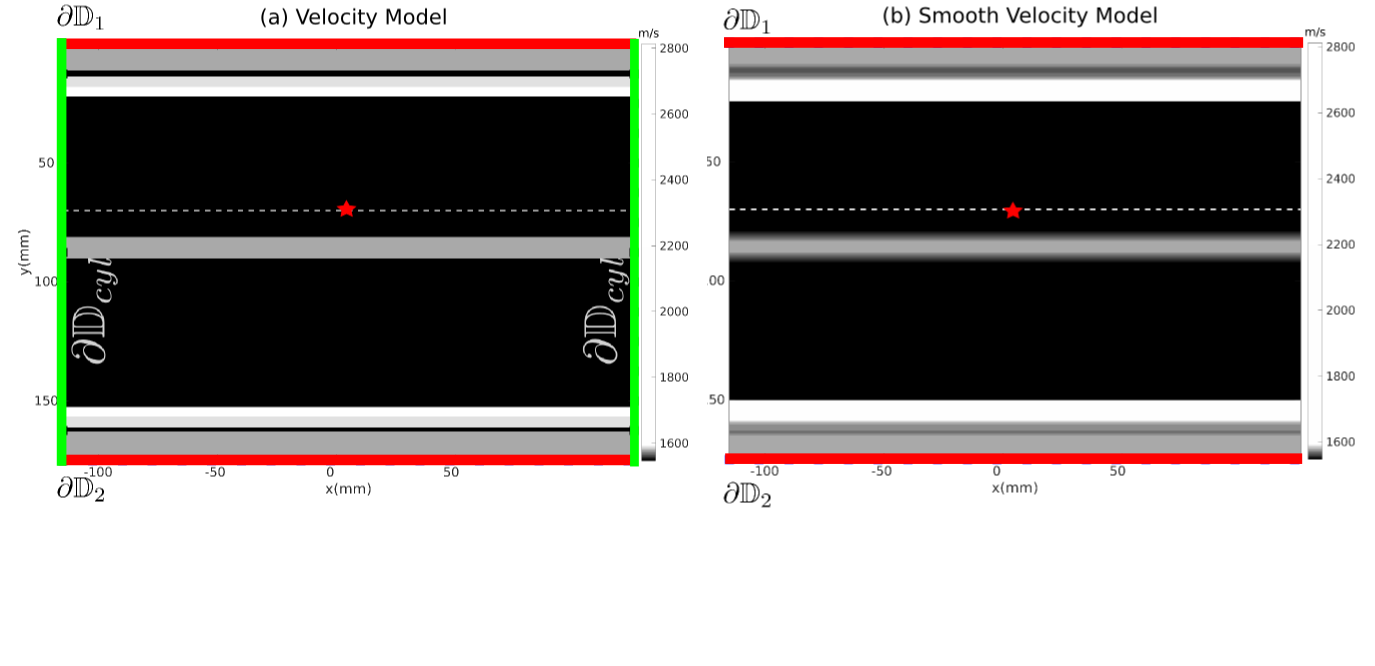}
\caption{(Color online) (a) True velocity model used in the first numerical experiment, corresponding to a 1.5D model associated with a cross-line of a human head model (see Fig. \ref{fig:brain} and Table \ref{table:brain}). The red
star and the dashed line represent the focal point and plane, respectively. 
For the Time-Reversal Focusing experiment associated with Eq. (\ref{eq:T2c}) (see the first column in Fig. \ref{fig:T1}), wavefields emanating from the focal point and recorded at evenly sampled receivers distributed along a closed boundary 
$\partial {\mathbb{D}}_1 \cup \partial {\mathbb{D}}_2  \cup \partial {\mathbb{D}}_{cyl} $ (thick red and green lines) are used.
For the Time-Reversal Focusing experiment associated with Eq. (\ref{eq:T2l}) (see the second column in Fig. \ref{fig:T1}), only wavefields recorded along horizontal
boundaries $\partial {\mathbb{D}}_1 \cup \partial {\mathbb{D}}_2  $
(thick red lines) are used.
For the focusing experiment associated with Eqs. (\ref{eq:T7v1}) and (\ref{eq:T9}) (see the third and fourth columns in Fig. \ref{fig:T1}), a total of evenly sampled 481$\times$2 co-located sources and receivers (indicated by the thick red lines) 
are used to compute reflection data along the upper ($\partial {\mathbb{D}}_1$) and the
lower ($\partial {\mathbb{D}}_2$) horizontal boundaries. 
Standard Marchenko methods are employed to retrieve focusing functions $f_1$ and $f_2$ using reflection data associated with $\partial {\mathbb{D}}_1$ and $\partial {\mathbb{D}}_2$, respectively \cite{wapenaar14}.
(b) Smooth velocity model used to compute the initial focusing function emanating from the focusing point (red star) and recorded along the upper ($\partial {\mathbb{D}}_1$) and the lower ($\partial {\mathbb{D}}_2$)
horizontal boundaries (thick red lines). 
}
  \label{fig:2D}
  \end{figure}

For illustration purposes, the right-hand sides of Eqs.  (\ref{eq:T2c}), (\ref{eq:T2l}), (\ref{eq:T7v1}) and (\ref{eq:T9}) are computed in a two-dimensional layered medium (Fig. \ref{fig:2D}(a)).
The focusing function $f_1$ is retrieved using a standard configuation \cite{Neut15,Thorbecke17}.
More precisely, iterative substitution of the coupled Marchenko equations allows to retrieve up- and down-going components of
focusing functions associated with arbitrary
locations in a medium. The methodology requires as input the single-sided reflection
response at the acquisition surface and an estimate of the initial focusing function, i.e. the Time-Reversed
direct wavefield from the specifed location in the subsurface to the acquisition surface. 
Here, to retrieve the focusing function $f_1$, reflection data  are then collected along the \textit{upper} boundary of the model ($\partial \mathbb{D}_1$ in Fig. \ref{fig:2D}(a)),
while the estimate of the initial focusing function
with a  $0.8$ MHz Ricker wavelet emanating from the focal point (red star in Fig. \ref{fig:2D}(b))
is computed in a smooth velocity model (see Fig. \ref{fig:2D}(b)).
Similarly, the focusing function $f_2$ is retrieved using reflection data collected along the \textit{lower} 
boundary of the model ($\partial \mathbb{D}_2$ in Fig. \ref{fig:2D}(a)). 
The estimate of the initial focusing function emanating from the focal point (red star in Fig. \ref{fig:2D}(b)) to the 
lower boundary receivers is also computed in the smooth velocity model in (Fig. \ref{fig:2D}(b)). 

Note that all data used in this paper are computed using a Finite Difference Time Domain vector-acoustic forward solver \cite{Thorbecke17}.

The solutions (i.e., the left-hand sides) from Eqs. (\ref{eq:T2c}), (\ref{eq:T2l}), and (\ref{eq:T7v1}) have infinite support in time, which could be disadvantageous for various applications. 
Things are different when Eq. (\ref{eq:T9}) is considered: since the focusing functions $f_1$ and $f_2$ are confined in time and space by 
the direct propagation path from the boundary to the focal point \cite{burridge80}, so is their superposition $f$. Hence, the solution associated with Eq. (\ref{eq:T9}) 
seems preferable for wavefield focusing in finite time rather than those related to Eqs. (\ref{eq:T2c}), (\ref{eq:T2l}), and (\ref{eq:T7v1}). 
More precisely, the real part of the focusing function $f$ contains a series of wavefronts that are emitted into the medium from the upper and lower boundaries, and only the first of these 
wavefronts reaches the focal point. 
The remaining events are encoded such that any ingoing reflection of the first wavefront is canceled.
The focusing conditions satisfied by Time-Reversed acoustics and Finite Time Focusing  differ drastically with respect to wavefield propagation in the focal plane.
While in Time-Reversed acoustics no constraint is posed on the propagation along the focal plane before or after time $t=0$, Finite Time Focusing 
limits the interaction of the wavefield with the focal \textit{plane} at the focal \textit{point} and at time $t=0$ only. \\

  \begin{figure}
  \centering
   \includegraphics[width=0.99\textwidth]{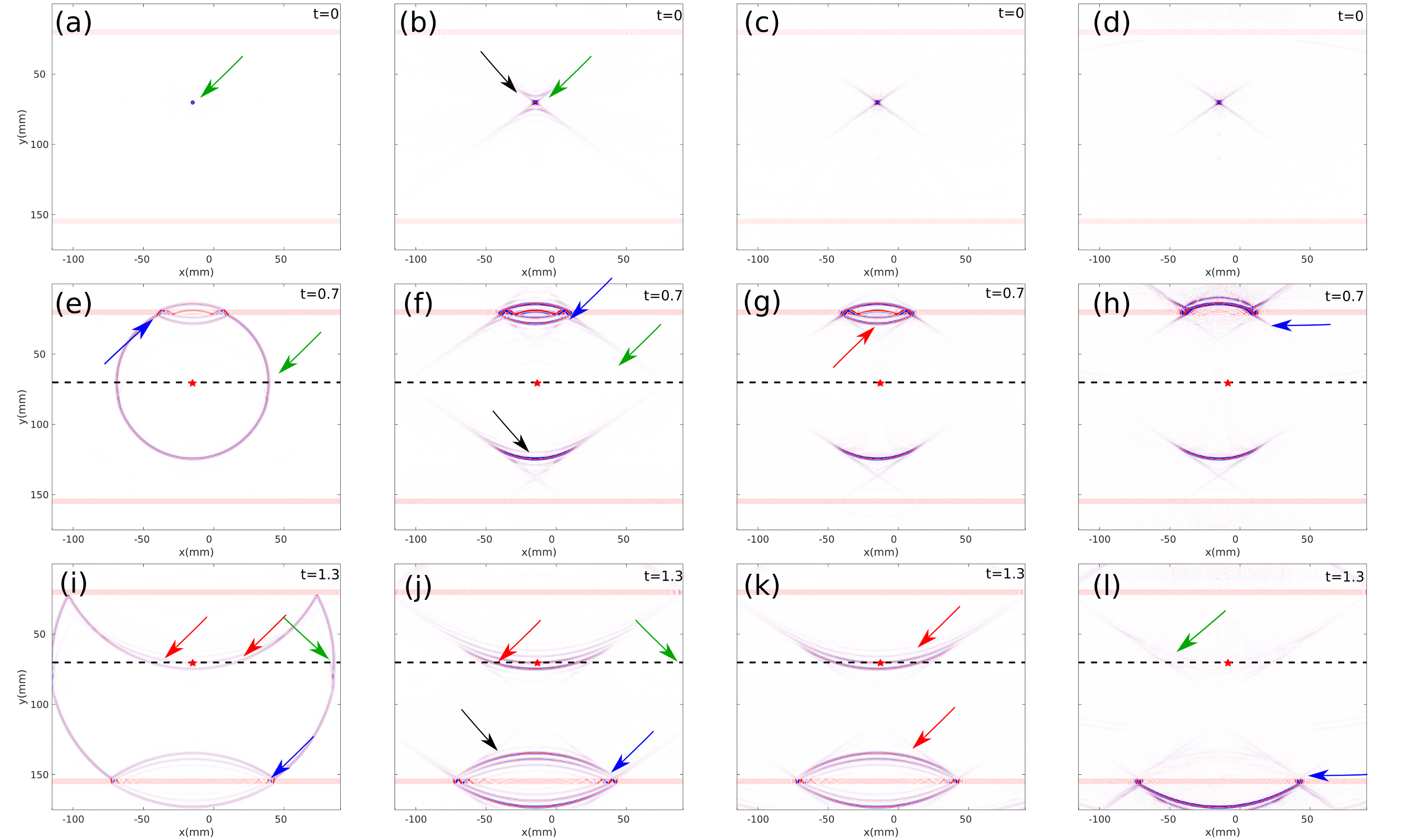}
\caption{(Color online) First Column: Snapshots of the Time-Reversed solution when a closed boundary is considered (Eq. (\ref{eq:T2c})). The focusing condition is satisfied, and
the wavefield at time $t=0$ is perfectly isotropic (green arrow). 
At time $t>0$ direct (green arrows) as well as scattered (blue arrows)
components of the wavefield are properly reconstructed. Red arrows indicate propagation of scattered waves through the focal plane. Light-red horizontal strips indicate strong reflectors, shown here for interpretation only,
while the red star and the black dashed line stand for the focal point and plane, respectively. 
Second Column: Snapshots of the Time-Reversed solution when partial boundaries are considered (Eq. (\ref{eq:T2l})). 
Due to the finite extent of the injection boundaries $\partial \mathbb{D}_1$ and $\partial \mathbb{D}_2$,
 the wavefield at time $t=0$ is not perfectly isotropic (green arrow), and artefacts, with maximum amplitude $\sim 5\%$ of the focus magnitude, contaminate the wavefield throughout the entire simulation (black arrows).
 At times $t>0$ scattered components of the wavefield are relatively well reconstructed (blue arrows), but the direct component of the wavefield exhibits distorted amplitudes along the horizontal direction (green arrows). 
 Red arrows indicate propagation of scattered waves through the focal plane.
 Third Column: Snapshots corresponding to Standard (double-sided) Marchenko Focusing (Eq. (\ref{eq:T7v1})). The focusing condition is only satisfied at time $t=0$ 
 At times $t>0$ scattered (red arrows) components of the wavefield are \textit{not} suppressed by destructive interference with propagation of the coda of $f$.
 Fourth Column: Snapshots corresponding to Finite Time Focusing (Eq. \ref{eq:T9}). The focusing condition is satisfied except for low amplitude artefacts, 
 with amplitude $\sim 2\%$ of the focus magnitude, propagating along the focal plane at times $t>0$ (green arrows).
 Note that the wavefield at time $t=0$ is not supposed to be vanishing throughout the domain (black arrows indicate propagation of the coda of $f$).
 At times $t>0$ scattered (blue arrows) components of the wavefield are suppressed by destructive interference with propagation of the coda of $f$.}
  \label{fig:T1}
  \end{figure}

We illustrate this in Fig. \ref{fig:T1} by showing propagation snapshots associated with the right-hand sides of Eqs.  (\ref{eq:T2c}), (\ref{eq:T2l}), (\ref{eq:T7v1}) and (\ref{eq:T9}).
Note that for the sake of brevity in the following we only focus on positive times, but identical considerations apply for the acausal components of the 
wavefields associated with Eqs. (\ref{eq:T2c}), (\ref{eq:T2l}), and (\ref{eq:T9}), while no acausal Green's functions terms propagate in Eq. (\ref{eq:T7v1}).
In Time-Reversed acoustics, the superposition of an acausal and a causal Green's function focusing and propagating away from ${\bf{x}} = {\bf{x}}_A$, is expected (Eqs. (\ref{eq:T2c}) and (\ref{eq:T2l})). 
Propagation around the foci is perfectly isotropic when Eq. (\ref{eq:T2c}) is used (green arrows in Figs. \ref{fig:T1}(a,e,i)), while the solution of Eq. (\ref{eq:T2l})
results in spurious events (black arrows in Fig. \ref{fig:T1}(b,f,j)) and artefacts, especially in the estimates of the direct wavefield along the focal plane 
(compare the amplitude of the wavefronts indicated by the green arrows in Figs.  \ref{fig:T1}(e,i) and \ref{fig:T1}(f,j)). 
These low amplitude artefacts are due to the finite extent of the horizontal boundaries employed in our numerical experiment when Eq. (\ref{eq:T2l}) is considered \cite{wapenaar14b}.
Note that in any case reflected waves propagating through the focal plane are well recovered both by Eqs. (\ref{eq:T2c}) and (\ref{eq:T2l}) (red arrows in Figs. \ref{fig:T1}(i) and \ref{fig:T1}(j)).
In Standard (double-sided) Marchenko Focusing (Eq. (\ref{eq:T7v1})), focusing is achieved at time $t=0$, but at later times Green's functions terms propagate within the layer embedding the focal plane 
(red arrows in Fig. \ref{fig:T1}(k)).
In Finite Time Focusing, destructive interference of up- and down-going wavefields prevents primary as well as multiple reflections to propagate through the focal plane at any time
(blue arrows in Fig. \ref{fig:T1}(h,l)).
The interaction of the wavefield with the layer embedding the focal point is therefore limited to the propagation of the direct components of $f$. 
Note that no direct or scattered waves propagating from and to the acquisition surfaces interact with the focal \textit{plane} except that at the
focal \textit{point}.\\



    \begin{figure}
  \centering
   \includegraphics[width=0.99\textwidth]{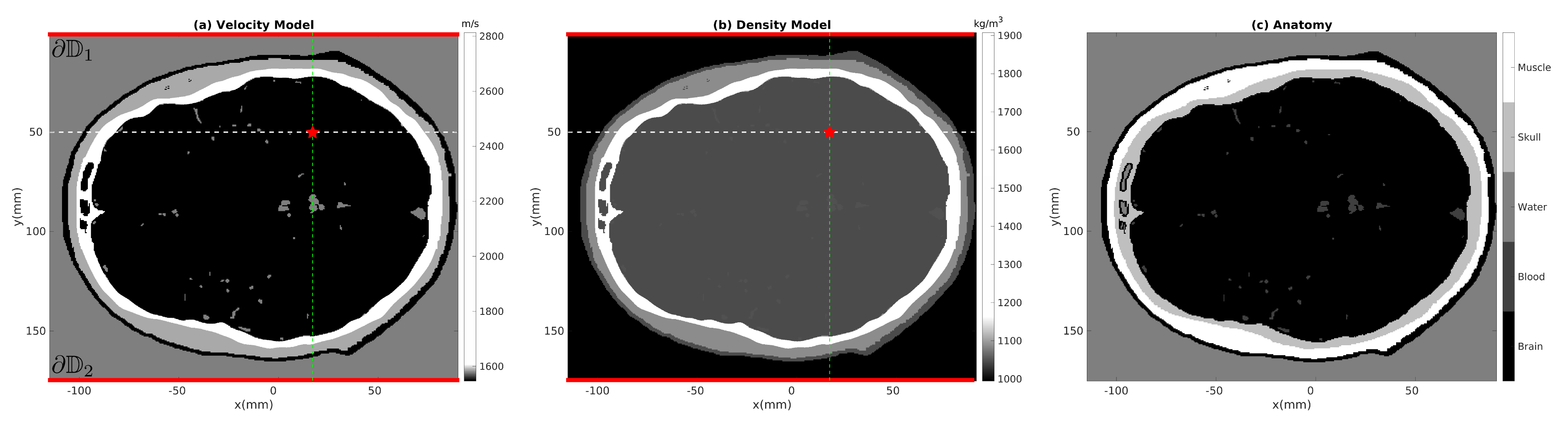}
\caption{(Color online) (a) True velocity model used in the second numerical experiment. The red star and the gray dashed line represent the focal point and plane, respectively. The green line indicates the 1D profile used for the first
numerical experiment.
For the Time-Reversal Focusing experiment associated with Eq. (\ref{eq:T2l}) (see first columns of Figs. \ref{fig:bT1} and \ref{fig:bT2}), wavefields
 emanating from the focal point and recorded at evenly spaced receivers located along horizontal
boundaries $\partial {\mathbb{D}}_1 \cup \partial {\mathbb{D}}_2  $
(thick red lines) are used.
For the focusing experiments associated with Eqs. (\ref{eq:T7v1}) and (\ref{eq:T9}) (see second and third columns of Figs. \ref{fig:bT1} and \ref{fig:bT2}), a total of 481$\times$2 evenly sampled co-located sources and receivers (thick red lines)
are used to compute reflection data along the upper ($\partial {\mathbb{D}}_1$) and the
lower ($\partial {\mathbb{D}}_2$) horizontal boundaries. 
Standard Marchenko methods are employed to retrieve focusing functions $f_1$ and $f_2$ using reflection data associated with $\partial {\mathbb{D}}_1$ and $\partial {\mathbb{D}}_2$, respectively.
This velocity model is also used to compute the initial focusing function emanating from the focal point (red star) and recorded along the upper ($\partial {\mathbb{D}}_1$) and the lower ($\partial {\mathbb{D}}_2$)
horizontal boundaries (thick red lines).
(b) True density model used in the second numerical experiments. 
(c) Anatomy of the brain used in the second numerical experiment. 
Keys as for (a).
}
  \label{fig:brain}
  \end{figure}
  
   \begin{table}
 \begin{center}
\begin{tabular}{ | m{5em} | m{2.8cm}| m{2.8cm} | } 
\hline
Tissue & velocity (m/s) & density (kg/m$^3$) \\ 
\hline
Muscle & 1588 & 1090 \\ 
\hline
Skull & 2813 & 1908 \\ 
\hline
Water & 1578 & 994 \\ 
\hline
Blood & 1578 & 1050 \\ 
\hline
Brain & 1546 & 1046 \\ 
\hline
\hline
\end{tabular}
\caption{Velocity and density values for the head model used in the second experiment (see Fig. \ref{fig:brain}).}
\label{table:brain}
\end{center}
\end{table}
  
The theory and methodology presented here hold  also for laterally variant models, and we show this by applying our focusing strategy to a second numerical experiment.
In this case we consider a model consisting of a slice of a human head  (see Fig. \ref{fig:brain} and Table  \ref{table:brain}) and explore the applicability of the method to medical imaging/treatment \cite{MIDA}.
This second example is chosen since it is particularly 
challenging for Marchenko focusing
due to the presence of thin layers, diffractors and dipping layers \cite{wapenaar14}.
As for the previous example, the focusing functions $f_1$ and $f_2$  are retrieved using standard Marchenko configurations, with reflection data collected along the upper 
and the lower boundaries of the model. Note that for actual therapy curved arrays are usually preferred over the linear acquisition configurations 
used here. The derivation of a new formulation of Finite Time Focusing to conform to more realistic therapeutical configurations will be the topic of future research.
Initial focusing functions with a  $0.8$ MHz Ricker wavelet emanating from the focal point (red star in Fig.  \ref{fig:brain}) to receivers at the upper and the 
lower boundaries are used. Note that for this example the initial focusing functions are computed in the true model (Fig. \ref{fig:brain}).\\

    \begin{figure}
  \centering
   \includegraphics[width=1\textwidth]{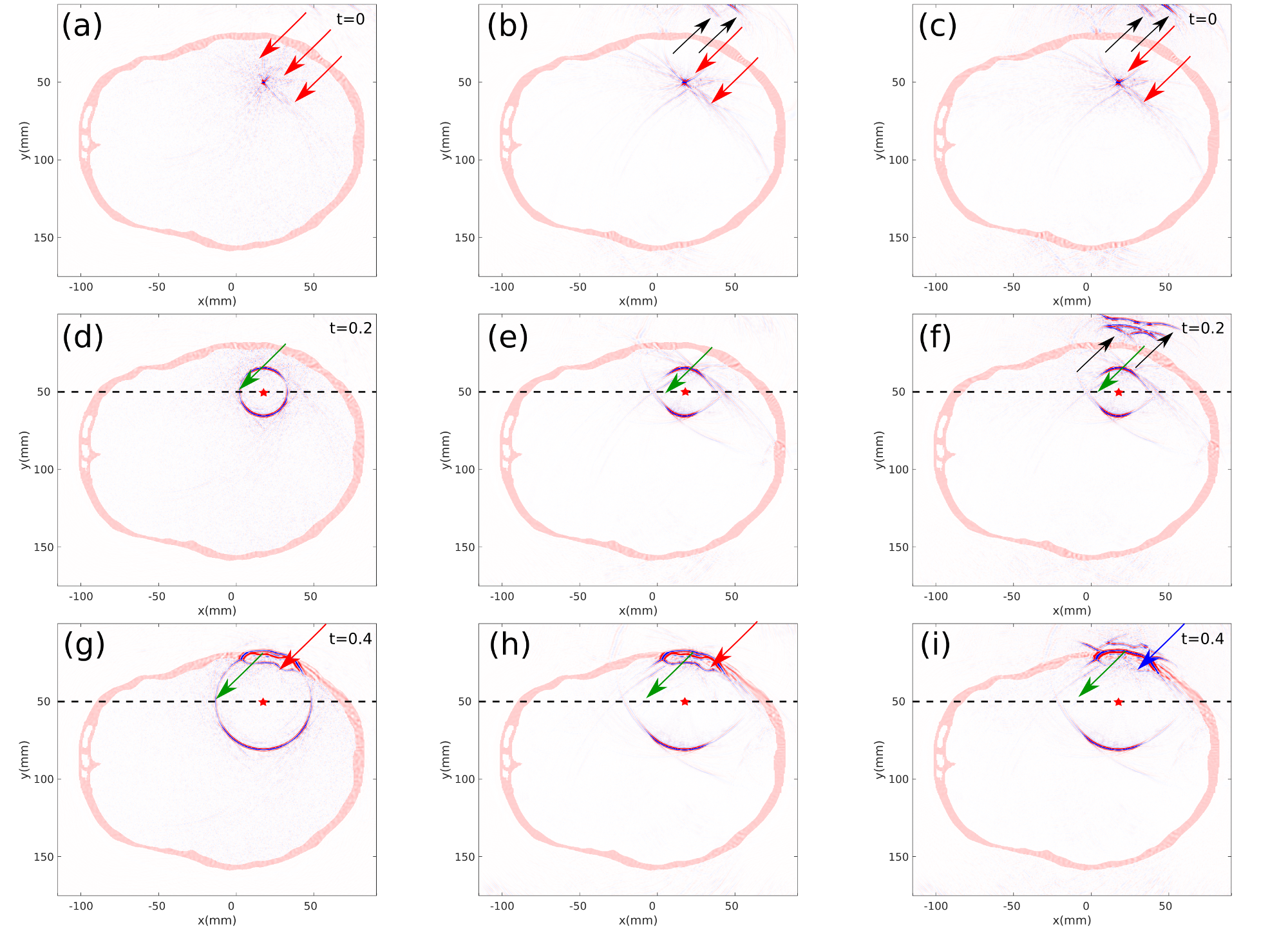}
\caption{(Color online) Focusing properties of  solutions of Eqs. (\ref{eq:T2l}), (\ref{eq:T7v1}) and (\ref{eq:T9}) in the time interval $\mbox{[0-0.4] s}$.  First column: Snapshots of the Time-Reversed solution 
when partial boundaries are considered (Eq. (\ref{eq:T2l})). Due to the finite extent of the injection boundaries $\partial \mathbb{D}_1$ and $\mathbb{D}_2$,
small amplitude artefacts contaminate the wavefield at time $t=0$ (red arrows in (a)).
Due to the strong lateral reflections, at times $t>0$ direct components of the wavefield are relatively well reconstructed (green arrows in (d) and (g)). 
The red arrow in (g) indicates a scattered wave reflected at the interface above the focal plane.
Second column: Snapshots  corresponding to Standard (double-sided) Marchenko Focusing (Eq. (\ref{eq:T7v1})). The focusing condition is satisfied except that for low amplitude artefacts, 
contaminating the domain at time $t=0$ (red arrow in (b)). Note that the wavefield at time $t=0$ is not supposed to be vanishing throughout the domain (black arrows indicate propagation of the coda of $f$).
At times $t>0$ scattered components of the wavefield are \textit{not} attenuated by destructive interference with propagation of the coda of $f$ (red arrow in (h)).
Third column: Snapshots of the focusing in finite time with minimal spatial exposure solution (Eq. (\ref{eq:T9})). The focusing condition is satisfied except for low amplitude artefacts, 
contaminating the domain at time $t=0$ (red arrow in (c)).
Note that the wavefield at time $t=0$ is not supposed to be vanishing throughout the domain (black arrows indicate propagation of the coda of $f$).
At times $t>0$ scattered components of the wavefield are attenuated by destructive interference with propagation of the coda of $f$ (blue arrow in (i)). Keys as in Fig. \ref{fig:T1}.
}.
  \label{fig:bT1}
  \end{figure}

    \begin{figure}
  \centering
   \includegraphics[width=1\textwidth]{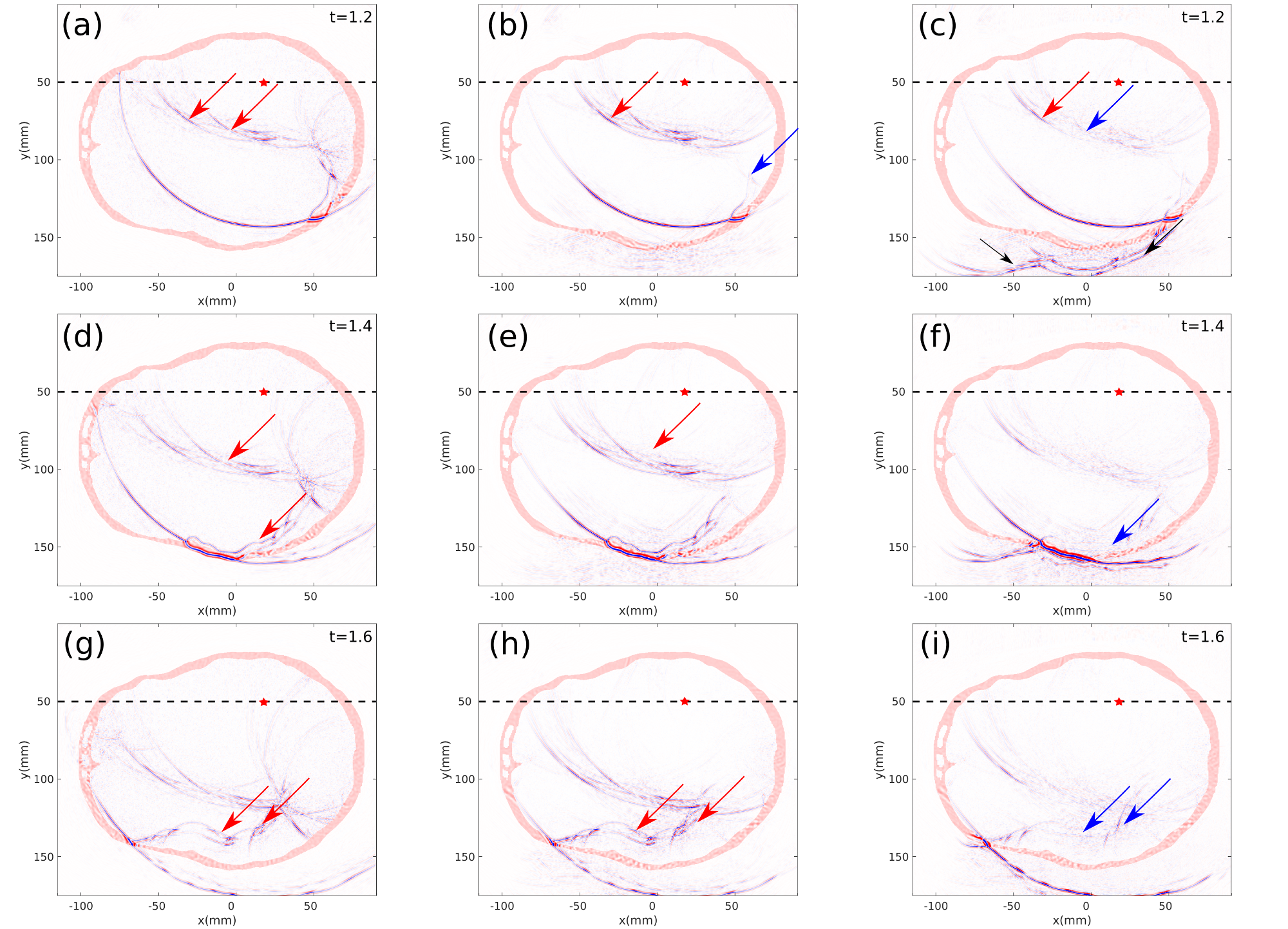}
\caption{(Color online) Focusing properties of  solutions of Eqs. (\ref{eq:T2l}), (\ref{eq:T7v1}) and (\ref{eq:T9}) in the time interval $\mbox{[1.2-1.6] s}$. First column: Snapshots of the Time-Reversed solution 
when partial boundaries are considered (Eq. (\ref{eq:T2l})). Red arrows point at reflections with the skull walls.
Second column: Snapshots corresponding to Standard (double-sided) Marchenko Focusing (Eq. (\ref{eq:T7v1})). The red arrows in (b, e, h) indicate scattered waves reflected at the interface above and below the focal plane.
Third column: Snapshots of the focusing in finite time with minimal spatial exposure solution (Eq. (\ref{eq:T9})). Black and blue arrows point at the coda of the focusing functions
and attenuated reflections, respectively. Keys as in Fig. \ref{fig:T1}.
}
   \label{fig:bT2}
  \end{figure}    

We first compare the focusing properties of  solutions of Eqs. (\ref{eq:T2l}), (\ref{eq:T7v1}) and (\ref{eq:T9}) by showing 
in Figs. \ref{fig:bT1} and \ref{fig:bT2} snapshots of the corresponding wavefields associated
with time intervals  $\mbox{[0-0.4] s.}$ and $\mbox{[1.2-1.6] s.}$, respectively.
Note that for the sake of brevity in the following we only focus on positive times, but identical considerations apply for the acausal components of the 
wavefields associated with Eqs. (\ref{eq:T2l}), and (\ref{eq:T9}), while no acausal Green's functions terms propagate in Eq. (\ref{eq:T7v1}).
In Time-Reversed acoustics (first column in Fig. \ref{fig:bT1}), the superposition of an acausal and a causal Green's function focusing and propagating away from ${\bf{x}} = {\bf{x}}_A$, is expected. 
However, due to the employed truncated boundaries, low amplitude artefacts occurring at time $t=0$ contaminate
the wavefield throughout the domain, especially in the proximity of the focal point (red arrows in Fig. \ref{fig:bT1}(a)). 
Similar artefacts at time $t=0$ also contaminate the wavefield associated with Eqs. (\ref{eq:T7v1}) (second column in Fig  \ref{fig:bT1}) and \ref{eq:T9} (third column in Fig  \ref{fig:bT1}).
In Figs. \ref{fig:bT1}(d) and \ref{fig:bT1}(g)  the wavefield associated with Eq. (\ref{eq:T2l}) is shown to propagate almost isotropically around the focal point.
More precisely, direct components of the wavefield $G(x_B,x_A)$, associated via Eq. (\ref{eq:T2l}) with laterally scattered waves $G(x,x_A)$ and $G(x,x_B)$  \cite{snieder2006}, 
 interact with the focal plane (green arrow in Fig. \ref{fig:bT1}(d)) at positive times. 
By contrast, the wavefields associated with Eqs. (\ref{eq:T7v1}) and (\ref{eq:T9}) do not exhibit similar components (green arrows in Figs. \ref{fig:bT1}(e,f,h,i)).
The red arrow in Fig. \ref{fig:bT1}(g) indicates a primary reflection associated with the wall of the skull above the focal plane. 
A similar event, corresponding to a Green's function term, is present  Fig. \ref{fig:bT1}(h).
On the other hand, the coda of the focusing function 
(black arrows in Figs. \ref{fig:bT1}(f)) interferes destructively with this reflection (blue arrow in Fig. \ref{fig:bT1}(i)).
Due to the complexity of the model, i.e.,
the presence of thin layers, diffractors and dipping layers \cite{wapenaar14},
the cancellation of the ingoing reflection is not perfect (red arrows in Fig. \ref{fig:bT2}(c)), but the amplitude of the reflected wave is generally \textit{reduced} (blue arrow in Fig. \ref{fig:bT2}(c)).
Similar considerations apply also for the reflection associated with the wall of the skull below the focal plane, where again the coda of the focusing function (black arrows in Fig. \ref{fig:bT2}(c)) 
is shown to interfere destructively (blue arrows in Figs. \ref{fig:bT2}(f) and \ref{fig:bT2}(i)) with the ingoing-reflection (red arrows in Figs. \ref{fig:bT2}(g) and \ref{fig:bT2}(h)).\\

    \begin{figure}
  \centering
   \includegraphics[width=0.99\textwidth]{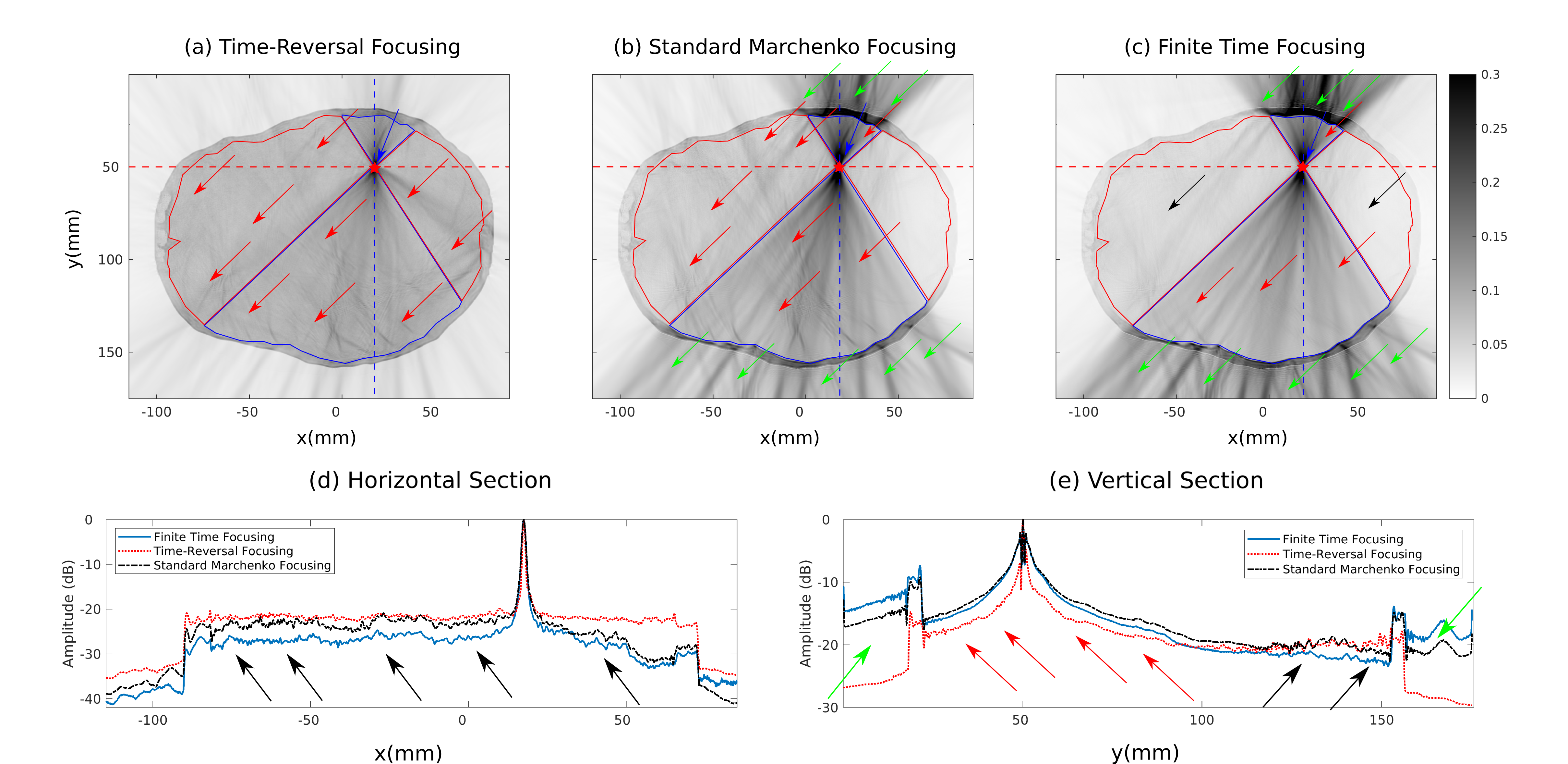}
\caption{(Color online) Normalized $L_2$ norm of the pressure wavefields associated with the left-hand sides of Eqs.  (\ref{eq:T2l}) (a), (\ref{eq:T7v1}) (b) and (\ref{eq:T9}) (c), respectively, plotted as functions of space.
In Standard Time-Reversal Focusing (a), the norm of the pressure wavefield exhibits a peak at the focal point (blue arrow in a), and significant values are almost
homogeneously distributed throughout the model (red arrows in (a)).
A similar distribution, with large values along the focal plane, is obtained when Standard (double-sided) Marchenko Focusing is used (b). 
In Finite Time Focusing, the wavefield is still exhibiting a peak at the focal point (blue arrow in Fig. (c)) while being somehow confined into a 
double cone centered at the focal point (blue cones in (c)).
Black and green arrows point at regions of the brain with minimal wavefield propagation and large
amplitude spots associated with the propagation of the coda of the focusing functions, respectively. Red and blue dashed lines indicate horizontal and vertical sections used in (d-e), respectively.
Horizontal (d) and vertical (e) slices of the maps in Fig. (a-c), plotted in Decibel scale ($20log_{10}(\lVert p \lVert)$).
Black arrows in (d) indicate large portions of the focal plane (red dashed lines in (a-c)) where wavefield propagation in Finite Time Focusing is significantly reduced as opposed to Time-Reversal 
and Standard (double-sided) Marchenko Focusing.
The red and black arrows in (e) indicate zones along the green dashed lines in  Fig. (a-c) where Finite Time Focusing and Time-Reversal Focusing 
involves slightly larger and slightly smaller wavefield intensity, respectively. Green arrows point at zones outside of the skull
where Standard (double-sided) Marchenko and Finite Time Focusing involve propagation of coda exhibiting large amplitudes (see green arrows in Fig. (c)).
Keys as in Fig. \ref{fig:brain}.
}
  \label{fig:maps}
  \end{figure}    


The differences between the three discussed focusing strategies  are visualized in another way  in Fig. \ref{fig:maps}, where the $L_2$ norm of
the pressure wavefields associated with Eqs.  (\ref{eq:T2l}), (\ref{eq:T7v1}) and (\ref{eq:T9}) is plotted as a function of space.
Note that all maps are normalized to allow proper comparison of the three focusing methods.
In Standard Time-Reversal Focusing, the  norm of the pressure wavefield exhibits a peak at the focal point (blue arrow in Fig. \ref{fig:maps}a), and significant values are almost
homogeneously distributed throughout the brain (red arrows in Fig. \ref{fig:maps}(a)). This indicates that wave propagation occurs in the entire brain, which could be undesirable for 
medical treatments designed to target the focal point while not affecting other portions of the brain.
Significant wavefield propagation throughout the brain occurs also when Standard (double-sided) Marchenko Focusing is employed (red arrows in Fig. \ref{fig:maps}(b)).
The situation is rather different when focusing is achieved via solution of Eq. (\ref{eq:T9}). Due to the peculiar focusing condition associated with Marchenko schemes \cite{wapenaar14c},
the corresponding wavefield still exhibits a peak at the focal point (blue arrow in Fig. \ref{fig:maps}(c)) while being mostly confined into a double cone 
centered at the focal point (blue cones in Fig. \ref{fig:maps}(c)).
Black and green arrows point at regions of the brain with minimal wavefield propagation inside the brain and large amplitude spots outside the brain associated 
with the propagation of the coda of the focusing functions, respectively.
The different performances of Time-Reversal, Standard (double-sided) Marchenko and Finite Time Focusing can be better appreciated in Figs. \ref{fig:maps}(d-e), where horizontal (d) and vertical (e) sections  of
the maps in Fig. \ref{fig:maps}(a-c) are plotted in Decibel scale ($20log_{10}(\lVert p \lVert)$). As expected, along the horizontal section (d) Finite Time Focusing exhibits reduced wavefield propagation, 
whereas along the vertical direction (e) the three diagrams are rather similar. 
Note that in Time-Reversal Mirroring wavefield propagation across the focal plane occurs before and after time $t=0$, in Standard (double-sided) Marchenko Focusing at time $t\geq0$ and in Finite Time Focusing the 
interaction of the wavefield with the focal point theoretically takes place
only at time $t=0$. Therefore, in Time-Reversal Mirroring and Standard (double-sided) Marchenko Focusing the norm of the wavefield at the focal point is intrinsically associated with both direct and scattered waves,
while in Finite Time Focusing it is theoretically only associated with direct components of the focusing function $f$. 
The overall focusing performances of the discussed methods are summarized in Table \ref{table:brain_performances}. 
The brain is divided in four domains, enclosed by the blue and the red curves in Figures \ref{fig:maps}(a-c), which represent
cones converging to the focal plane from the horizontal (i.e. the acquisition surface) and the vertical sides of the model, respectively. The norm of the wavefields associated with the three focusing strategies 
discussed in this paper is computed in the whole brain and in the areas enclosed by the blue and red curves.
Values are normalized with respect to the norms associated with Time-Reversal Mirroring in each individual domain. While in the whole brain and in the blue areas the three focusing strategies exhibit similar norm values, 
in the red areas
Finite Time Focusing involves significantly smaller values than Time-Reversal Mirroring and Standard (double-sided) Marchenko Focusing.


 \begin{table}
 \begin{center}
\begin{tabular}{ | m{5em} | m{2.8cm}| m{2.8cm} | m{2.8cm} |} 
\hline
 & Brain  & Blue Cones & Red Cones \\ 
\hline
SMF & +1\% & +16\% & -26\% \\ 
\hline
FTF & -14\% & +5\% & -45\% \\ 
\hline
\end{tabular}
\caption{Norm differences of the wavefields associated with the two new focusing strategies 
discussed in this paper (Standard (double-sided) Marchenko Focusing, here SMF, and Finite Time Focusing, here FTF) in the whole brain, first column, in the blue cones, second column, and
in the red cones, third column. Values are compared to the norm associated with Time-Reversal Mirroring in each domain.}
\label{table:brain_performances}
\end{center}
\end{table}

\section{DISCUSSION}

The wavefields resulting from the Time-Reversal and Standard (double-sided) Marchenko methods, as formulated by  Eqs. (\ref{eq:T2c}), (\ref{eq:T2l}) and (\ref{eq:T7v1}) have infinite support in 
time, which could be disadvantageous for various applications.
Things are different in Finite Time Focusing (Eq. (\ref{eq:T9})), which involves wavefields that are confined in time and space by the direct propagation path from the boundary to the focal point.
As can be observed in Figs.  \ref{fig:T1}, \ref{fig:bT1}, and \ref{fig:bT2}, the real part of the focusing function $f$ contains 
a series of wavefronts that once emitted into
the medium from the surrounding boundary interfere destructively with any ingoing reflection of the first pulse. Even when perfect focusing is not 
achieved, the amplitude of ingoing reflections is at least suppressed. 
Hence, the focusing function might be an attractive solution of the wave equation for focusing below strong acoustic contrasts. 
By canceling or reducing the amplitude of ingoing reflections,
we achieve the desirable situation of a single wavefront or reduced energy to reach the focal point and propagate along the focal plane.
Moreover, the peculiar nature of the focusing achieved by Eq. (\ref{eq:T9}) minimizes the spatial exposure to the incident wavefield of the layer embedding the focal point, and this could possibly be beneficial 
for sensitivity analysis and/or safety concern in medical treatment \cite{hughes17}.
Focusing functions associated with Eq. (\ref{eq:T9}) may also therefore be useful input for inversion. Akin to Green's functions, they obey the wave equation, 
which can be inverted for the medium properties $c\left( {\bf{x}}\right)$ and $\rho \left( {\bf{x}}\right)$. In particular cases, they may be preferred over Green's 
functions for this purpose, since the entire signals can be captured by a concise recording in the time domain and exhibit peculiar sensitivity distributions.
In the numerical tests considered here, we used either kinematically equivalent (first numerical experiment) or exact velocity models (second numerical experiment) to compute the initial focusing functions.
When a poor background model is used, solutions from above and below could focus at different points, and the terms associated with the Green's functions
in Eqs. (\ref{eq:T5v1})-(\ref{eq:T6v1}) and (\ref{eq:T5v2})-(\ref{eq:T6v2}) would not cancel out, thus violating the focusing condition exhibited by $f$.
Note that this restriction holds  also for the Time-Reversal method when applied from two sides.
The human skull involves some of the most critical challenges for Marchenko applications, i.e. the presence of thin layers, diffractors, dipping layers and strong absorption.
In our numerical test an acoustic and loseless model was employed.
Note that using a lossless head model allowed us to test the method on a simplified and yet very challenging problem.
However, neglecting dissipation, which plays a key role in medical treatment, limits the immediate applicability of the current algorithm 
of Finite Time Focusing, and a new theoretical framework to include absorption needs to be devised.
Recent research has shown that when media are accessible from two sides (which is a strict requirement in the focusing strategy discussed in this paper), 
Marchenko redatuming can be adapted to account for dissipation \cite{slob16}, and these
insights could foster future research devoted to extension of the proposed method to account for dissipative media.


\section{CONCLUSIONS}

A new integral representation has been derived for wavefield focusing in an acoustic medium. 
Unlike in the classical representation for this problem based on Time-Reversed acoustics, the input and 
output signals for this type of focusing are finite in time and only involve propagation of direct waves in the layer that embeds the focal point. 
This leads to a reduction of spatial and temporal exposure when wavefield focusing is applied in practice. The method has been validated numerically
for a head model consisting of hard (skull) and soft (brain) tissue. There results confirm that the proposed method can outperform classical Time-Reversed acoustics.

\section{ACKNOWLEDGMENTS}

This work is partly funded by the European Research Council (ERC) under the European Union's Horizon 2020 research and innovation programme (grant agreement No: 742703). Joost van der Neut is grateful to Niels Grobbe (University of Hawaii) for stimulating discussions and for conducting some of the initial research that evolved into this contribution.

\bibliographystyle{apalike}
\bibliography{paper}
\end{document}